# Broad learning system with robust adaptive kernel

Haiquan Zhao[a] , Jinhui Hu[a], and Xin Lu[a]

[a]*School of Electrical Engineering, Southwest Jiaotong University, Chengdu 611756, China*

ABSTRACT

For the performance degradation problem of broad learning system (BLS) in non-Gaussian noise environment, the variant of BLS based on M-estimator shows good robust performance. However, in most cases, the determination of the optimal loss function is often very time-consuming due to the lack of prior knowledge of the sample data. Therefore, this paper constructs a variant of BLS based on adaptive robust kernel (AR-BLS) to improve the generalization performance of the model in non-Gaussian noise environment. Adaptive robust kernel function is a general loss function that includes many common M-estimator paradigms. By alternately optimizing model weights and adaptive robust kernel parameters, AR-BLS realizes the adaptive adjustment of model robustness under different outlier noise distributions without human intervention. In addition, the iterative convergence of AR-BLS algorithm is proved based on Zangwill's global convergence theorem. Simulation experiments on multiple public datasets and actual application scenarios verify the effectiveness of the proposed method.

## 1. Introduction

Broad learning system (BLS)[1] has been widely used in pattern recognition[2]-[5], fault detection[6]-[9], sentiment analysis[10]-[13], time series prediction[14]-[17] and other fields[18]-[23] due to its lightweight network structure and competitive training efficiency. However, the objective function of the traditional BLS is constructed based on the minimum mean square error (MSE) criterion, and its theoretical framework is based on the prior assumption that the error term follows a Gaussian distribution. In practical application scenarios, the probability density function of data error often cannot be determined in advance, which leads to the performance loss of BLS in a non-noisy environment.

In response, scholars have redesigned the objective function of BLS and proposed some robust BLS variants. Jin et al.[24] constructed a sparse robust variant of BLS by introducing the L1 norm into the objective function by making an independent assumption on the distribution law of training errors and system weights. This variant uses L1 norm to constrain the training error, which improves the robustness of the model to Laplacian distributed noise. However, the training error in the actual situation does not necessarily conform to the Gaussian or Laplacian distribution. In order to further enhance the generalization ability of the model, Liu et al.[25] used the Cauchy loss function to replace the least squares metric in the traditional BLS, and proposed a Cauchy regularization variant of BLS. The nonlinear form of the Cauchy loss function makes it grow slowly when the error is large, thus reducing the impact of the noise term on the overall loss, which enables the model to handle data corrupted by Gaussian noise and outliers with different noise levels. In fact, the Cauchy loss function can be regarded as a special form of the M-estimator function, based on which Guo et al.[26] proposed a richer robust variant of BLS. M-estimator determines the contribution of each sample data to the model according to its influence function. Since these influence function values are small, the influence of outliers on the model is effectively weakened, and the robustness of the model to outlier noise is achieved.

Compared with the traditional BLS, the above variants all show better robustness in the non-Gaussian noise environment. However, the choice of loss function is often problem specific due to the variety of M-estimator functions and the fact that different characteristic functions usually match the corresponding noise distribution. In practice, for noisy environments where prior knowledge is lacking, the specific loss function form is usually determined by trial and error, which is often very time-consuming. Moreover, once the form of the loss function is fixed, the generalization performance of the model will also be limited. The adaptive robust kernel function[27] is a generalization loss function proposed by Barron, which generalizes common robust loss functions including Cauchy, Geman-McClure, Welsch, and generalized Charbonnier, Huber. The shape of the robust kernel function can be flexibly controlled by adjusting the hyperparameters of the generalization loss function. Therefore, in this paper, the objective function of BLS is reconstructed based on the adaptive robust kernel, and a variant of BLS based on the adaptive robust kernel is designed by considering the selection of the loss function as part of the model optimization. The main contributions of this paper are briefly described below.

1. Aiming at the performance degradation of BLS in non-Gaussian noise environment, a variant of BLS based on adaptive robust kernel function is proposed. Compared with the existing BLS variants constructed based on the fixed M-estimator function, the proposed method has better generalization performance in the non-Gaussian noise environment.

2. For the proposed variant, the iterative reweighted least squares (IRLS) method is used to train the model weights. By alternately optimizing the weight and the hyperparameters of the generalization loss function, the adaptive adjustment of the model loss function is realized, which avoids time-consuming manual parameter adjustment. Moreover, the convergence of the algorithm is proved based on Zangwill's theorem.

* *Corresponding author.*
E-mail address: hqzhao_swjtu@126.com(H. Zhao), 17695794976@163.com (X. Lu)

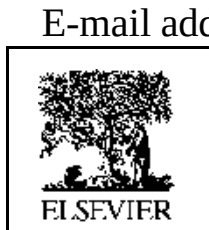

3. A large number of simulation experiments are carried out to test the performance of AR-BLS in multiple public datasets and practical application scenarios, and the experimental results verify the effectiveness of the proposed method.

The rest of this paper is structured as follows. Section II provides background on BLS and adaptive robust kernel functions. Section III describes the formula derivation process of AR-BLS as well as the corresponding parameter optimization method, and gives the proof of the convergence of the algorithm. In Section IV, experiments are carried out on University of California, Irvine (UCI) machine learning datasets and practical application scenarios to verify the performance of the proposed method. Finally, Section V gives a summary and induction.

For ease of presentation, Table 1 summarizes the main symbols used in this paper.

Tabel 1
The definition of notations

| Notation | Remark |
|---|---|
| BLS | Broad learning system |
| UCI | University of California, Irvine |
| IRLS | Iterative reweighted least squares |
| X | Input data matrix |
| A | The input matrix of the hidden layer |
| $\rho(e,\alpha,c)$ | Adaptive robust kernels |
| α | The shape parameter for adaptive robust kernel |
| c | The scale parameter for adaptive robust kernel |
| $P(e,\alpha,c)$ | Probability distribution function of the adaptive robust kernel |
| $Z(\alpha)$ | Partition function of the adaptive robust kernel |
| $\rho_\alpha(e,\alpha,c)$ | Negative log-likelihood function for the adaptive robust kernel |
| $\boldsymbol{\Lambda}$ | Adaptive robust kernel operator |
| $f(\bullet)$ | Algorithm iterative mapping |
| $\tilde{\alpha}$ | The candidate set of $\alpha$ |
| C | The parameter space of the algorithm |
| S | The solution set of the algorithm |
| RMSE | Root mean square error |
| MAE | Mean absolute error |

## 2. Related works

### 2.1 Broad Learning System

BLS is a new network architecture developed on the basis of random vector function link neural network (RVFLNN)[28]. As shown in Figure 1, the basic structure of BLS consists of three parts: input layer, hidden layer and output layer. Among them, the hidden layer design is the core of the BLS network structure, which is composed of two parts: feature nodes and enhancement nodes.

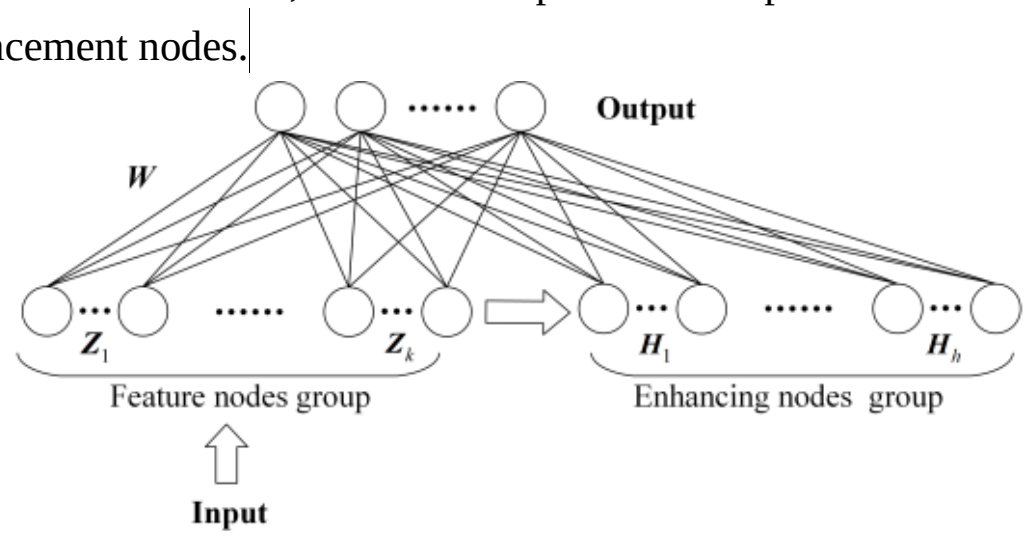

**Fig. 1** Basic architecture of BLS

For an input sample $\{\boldsymbol{X}, \boldsymbol{X} \in \boldsymbol{R}^{N\times t}\}$, where $N$ is the number of samples and $t$ is the sample feature dimension, feature nodes $\{\boldsymbol{Z}_i, i=1,2,\cdots n\}$ are generated by projecting the data into the feature space through the mapping function $f(\boldsymbol{X}W_{ei}+\boldsymbol{\beta}_{ei})$, and the set of top $n$ groups of feature nodes can be represented as $\boldsymbol{Z}^n=[\boldsymbol{Z}_1\cdots\boldsymbol{Z}_n]$. On this basis, the nonlinear activation function $\xi(\boldsymbol{Z}^n\boldsymbol{W}_{hj}+\boldsymbol{\beta}_{hj})$ is used to enhance the feature nodes to obtain the enhanced nodes $\{\boldsymbol{H}_j, j=1,2,\cdots m\}$, and the $m$ groups of enhanced nodes are concatenated into $\boldsymbol{H}^m=[\boldsymbol{H}_1\cdots\boldsymbol{H}_m]$. Where, $\boldsymbol{W}_{ei}$, $\boldsymbol{\beta}_{ei}$, $\boldsymbol{W}_{hj}$ and $\boldsymbol{\beta}_{hj}$ are all randomly generated and fixed. Combining $\boldsymbol{Z}^n$ with $\boldsymbol{H}^m$ yields an input vector $\boldsymbol{A}=[\boldsymbol{Z}^n \mid \boldsymbol{H}^m]$ containing information about all nodes. The output layer receives the merged vector from the hidden layer and trains the model weights by minimizing the error between the output and hidden layers, which can be expressed as follows.

$$\arg\min_{\boldsymbol{W}}\left(\|\boldsymbol{AW}-\boldsymbol{Y}\|_2^2+\lambda\|\boldsymbol{W}\|_2^2\right) \tag{1}$$

where $\boldsymbol{Y}$ is the label corresponding to $\boldsymbol{X}$; $\|\boldsymbol{AW}-\boldsymbol{Y}\|_2^2$ is used to control the minimization of training error, $\lambda\|\boldsymbol{W}\|_2^2$ is used to prevent model overfitting, $\lambda$ is the regularization coefficient.

Taking the derivative of Equation (2) with respect to $\boldsymbol{W}$ and setting it to zero, the weight expression can be expressed as follows.

$$\boldsymbol{W}=\left(\boldsymbol{A}^{\mathrm{T}}\boldsymbol{A}+\lambda\boldsymbol{I}\right)^{-1}\boldsymbol{A}^{\mathrm{T}}\boldsymbol{Y} \tag{2}$$

where $\boldsymbol{I}$ is the identity matrix and $\boldsymbol{A}^{\mathrm{T}}$ is the transpose of $\boldsymbol{A}$.

### 2.2 Adaptive robust kernel

Barron[27] proposed a loss function for a general framework, which is a superset of various classical robust losses, and its form can be expressed as follows.

$$\rho(e,\alpha,c)=\frac{|\alpha-2|}{\alpha}\left(\left(\frac{(e/c)^2}{|\alpha-2|}+1\right)^{\alpha/2}-1\right) \tag{3}$$

where $\alpha$ is a shape parameter that controls the robustness of $\rho(\bullet)$. $c$ is a scale parameter that controls the width of the quadratic region near the origin. The robustness properties of the general loss framework vary for different values of $\alpha$. By applying special treatment to the decodable singularities at $\alpha=0$ and $\alpha=2$ and as $\alpha\to0$, the final form of the general loss framework can be expressed as follows.

$$\rho(e,\alpha,c)=\begin{cases}\frac{1}{2}(e/c)^2 & \alpha=2\\ \log\left(\frac{1}{2}(e/c)^2+1\right) & \alpha=0\\ 1-\exp\left(-\frac{1}{2}(e/c)^2\right) & \alpha=-\infty\\ \frac{|\alpha-2|}{\alpha}\left(\left(\frac{(e/c)^2}{|\alpha-2|}+1\right)^{\alpha/2}-1\right) & \text{other situations}\end{cases} \tag{4}$$

It can be seen that the general loss framework covers Welsch, Geman-McClure, Cauchy, generalized Charbonnier, Huber as well as L2 loss functions. Figure 2 illustrates the curves of $\rho(e,\alpha,c)$ for different values of $\alpha$.

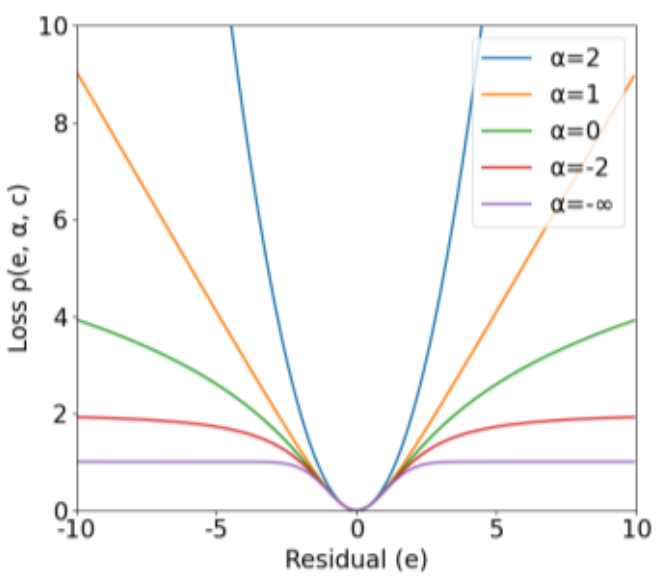

**Fig. 2** General loss function curves for different $\alpha$

In order to realize the adaptive adjustment of the general loss framework, the shape parameter $\alpha$ is used as an unknown parameter for joint optimization to minimize the $\rho(\bullet)$ :

$$\left(\boldsymbol{W}^*,\alpha^*\right)=\underset{(\boldsymbol{W},\alpha)}{\arg\min}\sum_{i=1}^{N}\rho\left(e_i\left(\boldsymbol{W},\alpha\right)\right) \tag{5}$$

For Equation (5), due to the monotonicity of $\rho(\bullet)$ with respect to $\alpha$, direct optimization may make $\alpha$ tend to extreme values (such as $\alpha\to 0$), resulting in all residuals being suppressed without discrimination. In this regard, by constructing a probability distribution $P(e,\alpha,c)$ based on $\rho(\bullet)$ :

$$P(e,\alpha,c)=\frac{1}{cZ(\alpha)}e^{-\rho(e,\alpha,1)} \tag{6}$$

where, $Z(\alpha)$ is the partition function that ensures the normalization of the probability distribution and is expressed as follows.

$$Z(\alpha)=\int_{-\infty}^{\infty}e^{-\rho(e,\alpha,c)}de \tag{7}$$

Therefore, the adaptive robust kernel function is defined as the negative log-likelihood of the distribution:

$$\rho_\alpha(e,\alpha,c)=-\log P(e,\alpha,c)=\rho(e,\alpha,c)+\log cZ(\alpha) \tag{8}$$

This transformation binds the loss function to a probability distribution, and the optimization of the parameter $\alpha$ will be constrained by the probability model. The regularization effect of the partition function forces the optimization process to choose an appropriate $\alpha$ instead of relying on manual parameter tuning. The curves of $\rho_\alpha(e,\alpha,c)$ for different $\alpha$ are shown in Figure 3.

**Fig. 3** Adaptive loss function curves for

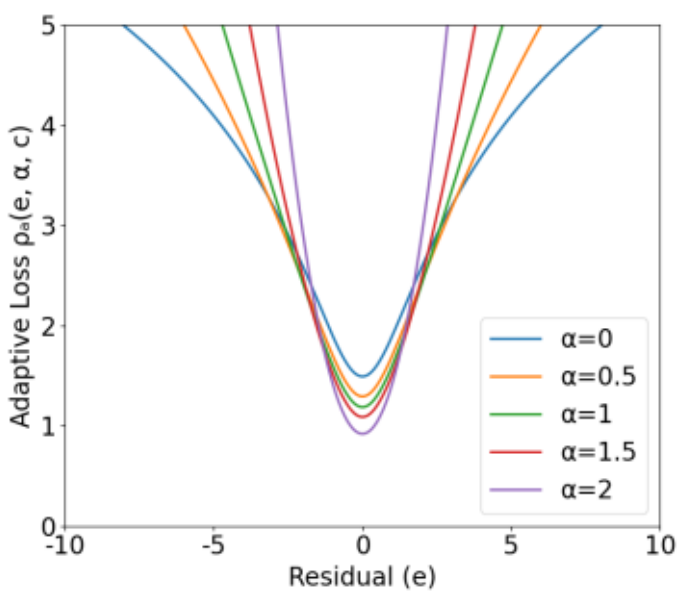

different $\alpha$

## 3. The AR-BLS algorithm

### 3.1 Algorithm derivation

By redesigning the objective function of BLS based on the adaptive robust kernel, the objective function of AR-BLS can be obtained as follows

$$J(\boldsymbol{W})=\underset{\boldsymbol{W}}{\arg\max}\sum_{i=1}^{N}\rho_\alpha(e_i,\alpha,c)-\frac{\lambda}{2}\|\boldsymbol{W}\|_2^2 \tag{9}$$

where $e_i=\boldsymbol{A}_i\boldsymbol{W}-\boldsymbol{Y}_i$, solving the gradient of $J(\boldsymbol{W})$ with respect to $\boldsymbol{W}$ yields:

$$\begin{aligned}\frac{\partial J(\boldsymbol{W})}{\partial \boldsymbol{W}}&=\sum_{i=1}^{N}\frac{\partial\rho_\alpha(e_i,\alpha,c)}{\partial e_i}\frac{\partial e_i}{\partial \boldsymbol{W}}+\lambda\boldsymbol{W}\\&=\sum_{i=1}^{N}-\varphi(e_i,\alpha,c)\boldsymbol{A}_i^T+\lambda\boldsymbol{W}\\&=\sum_{i=1}^{N}-\frac{\varphi(e_i,\alpha,c)}{e_i}e_i\boldsymbol{A}_i^T+\lambda\boldsymbol{W}\end{aligned} \tag{10}$$

where, $\varphi(e_i,\alpha,c)=\partial\rho_\alpha(e_i,\alpha,c)/\partial e_i$, can be expressed as follows.

$$\varphi(e,\alpha,c)=\begin{cases}\dfrac{e}{c^2} & \alpha=2\\ \dfrac{2e}{e^2+c^2} & \alpha=0\\ \dfrac{e}{c^2}\exp\left(-\dfrac{1}{2}(e/c)^2\right) & \alpha=-\infty\\ \dfrac{e}{c^2}\left(\dfrac{(e/c)^2}{|\alpha-2|}+1\right)^{\alpha/2-1} & \text{other situations}\end{cases} \tag{11}$$

The matrixization of Equation (10) can be obtained as follows.

$$\begin{aligned}\frac{\partial J(\boldsymbol{W})}{\partial \boldsymbol{W}}&=\sum_{i=1}^{N}-\boldsymbol{\Lambda}_i(\boldsymbol{Y}\boldsymbol{A}_i-\boldsymbol{W})\boldsymbol{A}_i^T+\lambda\boldsymbol{W}\\&=-\boldsymbol{A}^T\boldsymbol{\Lambda}(\boldsymbol{Y}-\boldsymbol{A}\boldsymbol{W})+\lambda\boldsymbol{W}\end{aligned} \tag{12}$$

where, $\Lambda_i=\varphi(u_i,\alpha,c)/u_i$, and $u_i$ denotes the normalized residual, which can be obtained from $u_i=0.6745\times(e_i/\text{med}(|e_i-\text{med}(e_i)|))$. Then, the adaptive robust kernel operator $\boldsymbol{\Lambda}$ is expressed as follows.

$$\boldsymbol{\Lambda}=\begin{pmatrix}\dfrac{\varphi(u_1,\alpha,c)}{u_1} & & \\ & \ddots & \\ & & \dfrac{\varphi(u_N,\alpha,c)}{u_N}\end{pmatrix} \tag{13}$$

Setting Equation (12) equal to zero yields the weight expression:

$$\boldsymbol{W}=\left(\boldsymbol{A}^T\boldsymbol{\Lambda}\boldsymbol{A}+\lambda\boldsymbol{I}\right)^{-1}\boldsymbol{A}^T\boldsymbol{\Lambda}\boldsymbol{Y} \tag{14}$$

According to Equation (14), the IRLS method is used to train the model weights.

$$\boldsymbol{W}^{t+1}=f\left(\boldsymbol{W}^t\right) \tag{15}$$

where $f(\boldsymbol{W})=\left(\boldsymbol{A}^T\boldsymbol{\Lambda}\boldsymbol{A}+\lambda\boldsymbol{I}\right)^{-1}\boldsymbol{A}^T\boldsymbol{\Lambda}\boldsymbol{Y}$. The specific process is shown in Algorithm 1.

Algorithm 1 AR-BLS

Input: training samples$\{X,Y\}$
1. Parameter setting: network parameters $n$, $q$, $m$, $p$, regularization parameter $\lambda$ , termination tolerance $\nu$ , maximum number of iterations $T$
2. Initialization: Set the initial weight $\boldsymbol{W}^0$ and build the input matrix $\boldsymbol{A}$
3. *for all* $t=1,2,\cdots,T$ *do*
4. Compute the $\boldsymbol{u}^t$ based on $\boldsymbol{W}^0$
5. Calculate the weighting matrix $\boldsymbol{\Lambda}$ according to Equation (13)
6. Update the $\boldsymbol{W}^t$ according to Equation (14)
7. Until $|\boldsymbol{W}^t-\boldsymbol{W}^{t+1}|_2^2<\nu$
8. end for
Output: $W$

### 3.2 Adaptive parameter optimization

Since the integral in the partition function $Z(\alpha)$ is unbounded for $\alpha<0$, the probability distribution $P(e,\alpha,c)$ is undefined. In this case, $\alpha$ is obtained by minimizing the negative log-likelihood of $P(e,\alpha,c)$, and the case $\alpha<0$ cannot be taken into account. Referring to [29], for $Z(\alpha)$ in Equation (7), an approximate partition function $\hat{Z}(\alpha)$ is constructed as follows.

$$\hat{Z}(\alpha)=\int_{-\varepsilon}^{\varepsilon}e^{-\rho(e,\alpha,1)}de \tag{16}$$

where, $\varepsilon$ is the truncated limit of the approximate integral. In this case, the partition function $\hat{Z}(\alpha)$ is integrated in a finite interval [ $-\varepsilon$ , $\varepsilon$ ], and the new negative log-likelihood function can be defined as follows.

$$\hat{\rho}_\alpha(e,\alpha,c)=-\log P(e,\alpha,c)=\rho(e,\alpha,c)+\log c\hat{Z}(\alpha) \tag{17}$$

Figure 4 shows the curve of $\hat{\rho}_\alpha(e,\alpha,c)$ under different $\alpha$. At this time, the loss function is still defined in the case of $\alpha$<0, and the optimization range of $\alpha$ is expanded.

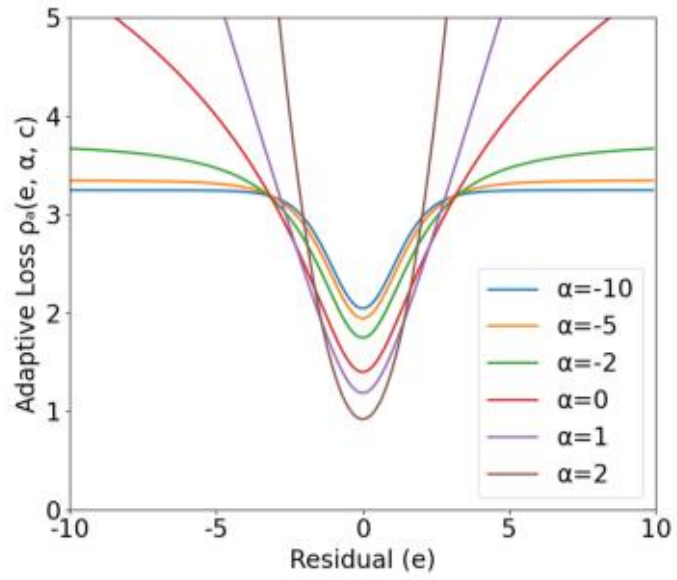

**Fig. 4** New adaptive loss function curves for different $\alpha$

It is worth noting that the new adaptive robust kernel function only optimizes the partition function $Z(\alpha)$, which is only related to $\alpha$ and independent of $\boldsymbol{W}$ in the formulation derivation of AR-BLS. Therefore, the new adaptive robust kernel will only affect the parameter optimization of $\alpha$, but will not affect Equation (14).

According to Equation (14), $\alpha$ is jointly optimized in the process of iterative reweight optimization. Calculate the current residual $\boldsymbol{u}$ based on the initial weight $W_0$, and estimate the value of $\alpha$ by minimizing Equation (16). This process can be achieved by performing one-dimensional grid search within [ $\alpha_{\min}$ ,2]. Once the maximum likelihood value of $\alpha$ is obtained, the model weights are calculated accordingly. The parameter optimization is achieved by alternately optimizing $\boldsymbol{W}$ and $\alpha$ in the iterative process. In a practical implementation, $\alpha$ can be set to -10, because for large residuals the difference between the weighting effect and $\alpha\to-\infty$ is small. For the parameter $c$, its value is usually determined based on the training residuals and kept fixed during the optimization process. The specific process is shown in Algorithm 2.

**Algorithm 2** Adaptive parameter optimization

**Input:** Initial weight $\boldsymbol{W}^0$
1. Set the candidate set of shape parameter $\tilde{\boldsymbol{\alpha}}\in[\alpha_{\min},2]$
2. *for all* $t$ =1,2, $\cdots$ ,$T$ *do*
3. Compute the $\boldsymbol{u}^t$ based on $\boldsymbol{W}^0$
4. Determine c in terms of $\boldsymbol{u}^t$
5. *for all* $\alpha$ *in* $\tilde{\boldsymbol{\alpha}}$ do
6. Minimize Equation (16) to determine $\alpha$
7. end for
8. Update the $\boldsymbol{W}^t$ according to Equation (13) and Equation (14)
9. end for

Output: $W$

3.3 Convergence analysis

For the weight optimization strategy adopted by AR-BLS, its iterative convergence can be proved by Zangwill's global convergence theorem[30], as shown in Theorem 3.1.

*Theorem 3.1*

For the iterative problem in the form of Equation (15) in Algorithm 1, $J(\bullet)$ is the objective function of the algorithm, $f(\bullet)$ is the iterative mapping of the algorithm, and let $\boldsymbol{C}$ be a compact set if the following conditions are satisfied:

Condition 1 Closed set constraint: All iteration points $\boldsymbol{W}^t$ and $\alpha^t$ belong to the closed set $\boldsymbol{C}$.

Condition 2 Property of descent: For any point $\boldsymbol{W}$ and $\alpha$ not in the solution set $\boldsymbol{S}$, the iterated map $f(\bullet)$ satisfies $J(f(\boldsymbol{W},\alpha))<J(\boldsymbol{W},\alpha)$.

Condition 3 Mapping closure property: The iterated map $f(\bullet)$ is closed on $\boldsymbol{C}\backslash\boldsymbol{S}$, that is, if $(\boldsymbol{W}^t,\alpha^t)\to(\boldsymbol{W}^*,\alpha^*)$ and $f(\boldsymbol{W}^*,\alpha^*)\to\gamma^*$, then $\gamma^*=f(\boldsymbol{W}^*,\alpha^*)$.

Then all solutions of the sequence { $\boldsymbol{W}^t,\alpha^t$ } generated by the algorithm belong to the solution set $\boldsymbol{S}$ and $f(\boldsymbol{W}^t,\alpha^t)$ converges to $f(\boldsymbol{S})$.

*Proof*

The objective function of AR-BLS can be simplified as follows

$$J(\boldsymbol{W},\alpha)=\underset{\boldsymbol{W}}{\arg\max}\sum_{i=1}^{N}\rho_\alpha(e_i,\alpha,c)+\lambda\|\boldsymbol{W}\|_2^2 \tag{18}$$

Define the solution set $S$ as the parameter pair ( $\boldsymbol{W}^*$ , $\alpha^*$ ) satisfying the following conditions:

$$\boldsymbol{S}=\left\{(\boldsymbol{W},\alpha)\middle|\nabla_{\boldsymbol{W}}J=0,\alpha=\underset{\alpha\in[-10,2]}{\arg\min}\,J(\boldsymbol{W},\alpha)\right\} \tag{19}$$

Equation (18) consists of a loss term $\sum_{i=1}^{N}\rho_\alpha(e_i,\alpha,c)$ and a regularization term $\lambda\|\boldsymbol{W}\|_2^2$. The adaptive robust kernel $\rho_\alpha(e_i,\alpha,c)$ in the loss term has $\rho_\alpha(e_i,\alpha,c)\geq0$ for any residual $e_i$. Combined with the nonnegativity of the regularization term ( $\lambda>0$ ), there exists a constant $\theta>0$ such that:

$$J(\boldsymbol{W},\alpha)\geq\theta \tag{20}$$

From the non-negativity of the loss term, it can be obtained that $J(\boldsymbol{W},\alpha)\geq\lambda\|\boldsymbol{W}\|_2^2$. Combined with Formula (20), it is not difficult to obtain:

$$\|\boldsymbol{W}\|\leq\sqrt{\frac{J(\boldsymbol{W},\alpha)}{\lambda}}\leq\sqrt{\frac{\theta}{\lambda}} \tag{21}$$

Thus, $\boldsymbol{W}$ belongs to the compact set $\boldsymbol{\omega}=\{W\,|\,\|W\|<\sqrt{\theta/\lambda}\}$. And since $\alpha\in[-10,2]$ is a closed interval, this interval is a compact set by the Heine-Borel theorem[31]. Then, the joint parameter space $\boldsymbol{C}=\boldsymbol{\omega}\times[-10,2]$ is a compact set and Condition 1 is proved.

The iterative process of AR-BLS is divided into two steps. First, $\boldsymbol{W}^t$ is fixed and $\alpha^{t+1}$ is optimized by minimizing the negative log-likelihood function via grid search. Then $\alpha^{t+1}$ is fixed and $\boldsymbol{W}^{t+1}$ is optimized by the IRLS method.

For fixed $\boldsymbol{W}^t$, the optimization problem reduces to the following.

$$\alpha^{t+1}=\underset{\alpha\in[-10,2]}{\arg\min}\,J\left(\boldsymbol{W}^t,\alpha\right) \tag{22}$$

Here all candidate points are directly compared by grid search on $\tilde{\boldsymbol{\alpha}}$, so for any $\boldsymbol{W}^{t+1}$

$$J\left(\boldsymbol{W}^t,\alpha^{t+1}\right)=\min_{\alpha\in\tilde{\boldsymbol{\alpha}}}J\left(\boldsymbol{W}^t,\alpha\right)\leq J\left(\boldsymbol{W}^t,\alpha^{t+1}\right) \tag{23}$$

For a fixed $\alpha^{t+1}$, the IRLS method updates the weights iteratively by a local quadratic approximation, and the second-order Taylor expansion of the adaptive robust kernel is $\rho_\alpha(e_i,\alpha,c)\approx\rho_\alpha\left(e_i^t,\alpha^{t+1},c\right)+\boldsymbol{W}_i^t\left(e_i-e_i^t\right)^2$, where

$$\boldsymbol{W}_i^t=\frac{1}{2}\frac{\partial^2\rho}{\partial e^2}\bigg|_{e=e_i^t} \tag{24}$$

After ignoring the constant term, the objective function is approximated as follows.

$$\tilde{J}(\boldsymbol{W})=\sum_{i=1}^{N}\boldsymbol{W}_i^te_i^2+\lambda\|\boldsymbol{W}\|_2^2 \tag{25}$$

Since $\tilde{J}(\boldsymbol{W})$ is a strongly convex function, its unique solution $\boldsymbol{W}^{t+1}$ satisfies

$$\tilde{J}\left(\boldsymbol{W}^{t+1}\right) < \tilde{J}\left(\boldsymbol{W}^{t}\right) \tag{26}$$

Combined with the convexity of $\rho_\alpha(e,\alpha,c)$ with fixed $\alpha$, the original objective function $J\left(\boldsymbol{W}^{t},\alpha^{t+1}\right)$ strictly decreases in the iterative reweighting step.

In summary, each alternating optimization iteration satisfies

$$J\left(\boldsymbol{W}^{t+1},\alpha^{t+1}\right) \le J\left(\boldsymbol{W}^{t},\alpha^{t}\right) \tag{27}$$

At this point, condition 2 has been proved.

The weights of AR-BLS are updated as follows

$$\boldsymbol{W}^{t+1} = \left(\boldsymbol{A}^T \boldsymbol{\Lambda}^t \boldsymbol{A} + \lambda \boldsymbol{I}\right)^{-1} \boldsymbol{A}^T \boldsymbol{\Lambda}^t \boldsymbol{Y} \tag{28}$$

Since $\boldsymbol{\Lambda}^t$ is continuously dependent on $\boldsymbol{W}^t$ and the matrix inversion is continuous under non-singular conditions, the IRLS mapping is closed on the compact set $\boldsymbol{C}$.

As shown in Equation (22), the update of $\alpha$ is achieved through grid search, and the candidate set is a finite discrete point set. For any convergent subsequence $\tilde{\boldsymbol{\alpha}}$, the limit point $\alpha^*$ must belong to the candidate set, and thus the mapping is closed. At this point, condition 3 is proved.

According to the three conditions of Zangwill's theorem, any limit point of the sequence $(\boldsymbol{W}^t,\alpha^t)$ generated by AR-BLS algorithm belongs to the solution set $\boldsymbol{S}$, that is, the algorithm converges to a stable point globally.

## 4. Experiments

In this section, simulation experiments are conducted on the UCI regression datasets and real-world industrial application scenarios. In addition to the traditional BLS, M-BLS constructed based on different M-estimator functions will be used to compare the performance with the proposed variant. During the experiment, the root mean square error (RMSE) and mean absolute error (MAE) are selected as the key criteria to measure the performance, which are defined as follows.

$$\text{RMSE} = \sqrt{\frac{1}{N}\sum_{n=1}^{N}\left(y_n - \hat{y}_n\right)^2} \tag{29}$$

$$\text{MAE} = \frac{1}{N}\sum_{n=1}^{N}\left|y_n - \hat{y}_n\right| \tag{30}$$

where $N$ represents the number of samples, $y_n$ is the true value of the $n$-th sample, and $\hat{y}_n$ is the predicted value of the $n$-th sample. In model performance evaluation, both MAE and RMSE are used to measure the deviation between the predicted value and the true value. MAE takes the average absolute error, which reflects the absolute magnitude of the overall deviation. RMSE is more sensitive to large errors and can highlight the shortcomings of the model in extreme predictions. Generally, smaller values of RMSE and MAE indicate better model performance.

In order to systematically evaluate the performance of BLS and its various variants, the experimental parameters are uniformly set in this section. For the traditional BLS, the grid search method is used to confirm its optimal network structure parameters. Specifically, the search range of the number of feature node groups $n$ is set as [1,20], and the search step size is 1. The search range for the number of feature nodes $q$ in each group is [1,20] with a search step size of 2. The number of enhanced node groups $m$ is fixed as 1. The number of enhanced nodes $p$ in each group is searched over a range of [1,200] with a search step size of 5[21]. Considering that the network structure of each BLS variant is unchanged, the network structure parameters of each variant in the experiment are set as the traditional BLS by default, so as to highlight the performance differences of each robust strategy on the BLS framework. For M-BLS, BLS variants based on Huber, Cauchy and Welsch functions are constructed respectively. To ensure fair experiments, the regularization coefficient $\lambda$ is uniformly set to $2^{-30}$. To reduce the effect of randomness, each set of experiments was repeated ten times and the average value was taken as the final result. All experiments were performed on a PyCharm workbench equipped with Intel(R) Core(TM) i5-6200U CPU, 2.30 GHz and 8GB RAM.

### 4.1 Experiments on the UCI regression datasets

In this subsection, seven real regression datasets from medical, biological, and economic fields are selected from the UCI machine learning repository to test the performance of the proposed method. In the data preparation phase, each data set is divided into training set and test set with a ratio of 1:1. The key parameters of each dataset, such as feature dimension and data size, are detailed in Table 2. Considering that data scale differences may interfere with experimental results, all datasets are normalized, and the value range is uniformly mapped to the interval [0,1]. In order to simulate the polluted data regression task in practical applications, different proportions of outlier noise and $\alpha$ stable noise with impulse characteristics will be added to the data set in the subsequent experiments to test the robustness of the model in the simulated non-Gaussian noise environment.

Tabel 2
UCI regression datasets

| Dataset | Attributes | Training | Testing |
|---|---|---|---|
| Housing | 20 | 183 | 182 |
| Bodyfat | 14 | 126 | 126 |
| Clevend | 13 | 149 | 148 |
| Wine | 11 | 800 | 800 |
| Abalone | 7 | 712 | 712 |
| Slump | 7 | 52 | 52 |
| Strike | 6 | 313 | 312 |

Firstly, experiments are performed in different outlier noise level environments. Training samples are randomly selected according to different proportions $P$, and random values in the range [ $y_{\min}, y_{\max}$ ] are added to the corresponding output to construct contaminated datasets with different noise levels. Table 3 shows the test RMSE ± STD( $\times 10^{-2}$ ) of each variant on the corresponding dataset under different noise rate $P$. Among them, the best result is marked in bold. The corresponding training times are shown in Table 4. According to the experimental results in the table, it is not difficult to obtain:

(1) The performance degradation of conventional BLS based on L2 loss function is very obvious in outlier noise environment. As the noise rate increases, its test RMSE increases greatly. For large residuals generated by outliers, the squaring of the L2 loss magnifies this difference twice. This causes outliers to contribute far more than their number to the total loss, forcing the model to deviate from the fit of most normal data.

(2) The three M-BLS constructed based on different M-estimator functions show better performance than the basic BLS in most cases. At low noise rates, M-BLS(Huber) performs relatively better. Huber function has high sensitivity to small residuals, and can give larger weight to those samples that are close to the true value, so as to improve the fitting accuracy of the model to normal data. With the increase of noise rate, the performance advantage of M-BLS(Cauchy) is more obvious. For large residuals, the penalty of Cauchy function grows slowly and is able to reduce the contribution of outliers to the total loss, thereby mitigating their impact on the model. The performance of M-BLS(Welsch) is relatively balanced. Welsch function achieves interior point fitting and outlier suppression through exponential saturation mechanism, and balances the sensitivity to Gaussian noise and the robustness to heavy-tail noise of the model.

(3) AR-BLS achieves better performance than M-BLS constructed based on fixed-form M-estimator functions in most cases. Only on the

Slump dataset with $P$=10% and the Housing dataset with $P$=20%, its test RMSE is slightly higher than that of M-BLS(Welsch) and M-BLS(Cauchy), respectively, but still better than the other two variants. For different proportions of outlier noise, the adaptability of the fixed form M-estimator function is poor, and the test RMSE of M-BLS increases significantly more than AR-BLS. By dynamically adjusting the shape parameter $\alpha$ ($2 \rightarrow$-10), the adaptive kernel function can smooth the weight curve from quadratic growth to exponential saturation, so that the model maintains good robustness on datasets with different proportions of outliers. In addition, the fixed form M-estimator function handles outliers in a relatively fixed way. For example, Huber loss reduces the weight of outliers in a linear manner after the residual exceeds a threshold, Cauchy loss and Welsch loss reduce the weight according to their respective fixed functional form. However, the adaptive robust kernel function can adjust $\alpha$ in the continuous space according to the specific situation of the outliers, and obtain the loss function of any intermediate form, so as to find the optimal balance between fitting the interior point and suppressing the outliers. Benefiting from the characteristics of adaptive robust kernel, AR-BLS has significant advantages in the processing of outliers and model generalization ability.

(4) Compared with the basic BLS as well as the three M-BLS, AR-BLS usually takes longer training time, which is mainly because the parameters of its robust kernel function need some time to be adaptively adjusted. However, it should be pointed out that the loss of training time is controlled within an acceptable range compared to the performance improvement.

Tabel 3
Performance of different methods on regression datasets with different noise rates

| Dataset | Noise rate | The RMSE of test data (RMSE±STD $\times 10^{-2}$) | | | | |
|---|---|---|---|---|---|---|
| | | BLS | M-BLS(Huber) | M-BLS(Cauchy) | M-BLS(Welsch) | AR-BLS |
| Housing | $P$=10% | 15.12±2.32 | 14.13±3.23 | 14.26±3.41 | 14.50±3.36 | 13.76±3.26 |
| | $P$=20% | 16.91±3.11 | 14.59±2.58 | 14.62±2.76 | 14.94±2.82 | 14.64±2.71 |
| | $P$=30% | 19.14±4.01 | 15.36±2.75 | 15.32±2.82 | 16.04±3.41 | 15.31±2.90 |
| Bodyfat | $P$=10% | 19.43±3.41 | 18.02±3.27 | 17.99±3.28 | 18.10±3.23 | 17.41±2.83 |
| | $P$=20% | 23.47±5.67 | 21.35±4.03 | 21.25±3.75 | 21.38±3.80 | 20.94±3.74 |
| | $P$=30% | 26.78±6.06 | 23.73±5.02 | 23.38±4.86 | 24.64±5.39 | 22.06±3.52 |
| Clevend | $P$=10% | 24.74±1.22 | 24.30±1.70 | 24.19±1.51 | 24.20±1.18 | 23.81±1.02 |
| | $P$=20% | 27.48±2.31 | 26.10±2.73 | 25.99±2.63 | 26.25±2.39 | 25.58±2.03 |
| | $P$=30% | 30.67±4.43 | 28.53±3.44 | 28.41±3.45 | 29.20±4.09 | 27.68±2.19 |
| Wine | $P$=10% | 14.43±0.43 | 13.59±0.29 | 13.54±0.27 | 13.54±0.27 | 13.53±0.26 |
| | $P$=20% | 16.75±0.64 | 14.27±0.42 | 14.09±0.40 | 14.46±0.47 | 13.73±0.32 |
| | $P$=30% | 20.16±0.67 | 16.39±0.60 | 16.06±0.60 | 17.76±0.71 | 14.14±0.46 |
| Abalone | $P$=10% | 12.38±2.23 | 12.35±1.98 | 12.46±2.04 | 12.48±2.00 | 12.31±1.93 |
| | $P$=20% | 13.55±3.27 | 12.08±2.00 | 12.21±1.96 | 12.29±1.98 | 12.01±1.83 |
| | $P$=30% | 16.19±4.36 | 12.92±3.02 | 12.71±2.85 | 13.79±3.43 | 12.51±2.16 |
| Slump | $P$=10% | 15.26±2.52 | 13.49±4.45 | 13.28±4.38 | 13.20±4.31 | 13.25±4.28 |
| | $P$=20% | 16.91±2.75 | 15.33±5.13 | 14.81±4.63 | 14.75±4.55 | 14.75±4.48 |
| | $P$=30% | 19.03±2.79 | 22.69±9.83 | 22.25±10.09 | 18.94±2.99 | 17.28±7.12 |
| Strike | $P$=10% | 27.19±1.05 | 27.89±3.57 | 27.51±2.43 | 27.29±1.83 | 26.79±0.81 |
| | $P$=20% | 28.92±2.18 | 28.41±2.86 | 28.53±3.19 | 28.00±1.36 | 27.95±0.96 |
| | $P$=30% | 30.27±1.10 | 29.33±1.72 | 29.98±3.21 | 30.56±2.61 | 29.16±1.59 |

Tabel 4
Training time of different methods on the regression datasets

| Dataset | Noise rate | The training time of different methods (s) | | | | |
|---|---|---|---|---|---|---|
| | | BLS | M-BLS(Huber) | M-BLS(Cauchy) | M-BLS(Welsch) | AR-BLS |
| Housing | $P$=10% | 0.0743 | 0.2582 | 0.3026 | 0.2501 | 0.5728 |
| | $P$=20% | 0.0759 | 0.2609 | 0.2566 | 0.1907 | 0.5708 |
| | $P$=30% | 0.0757 | 0.2627 | 0.2843 | 0.2443 | 0.2378 |
| Bodyfat | $P$=10% | 0.0736 | 0.0961 | 0.0987 | 0.1010 | 0.4043 |
| | $P$=20% | 00466 | 0.0461 | 0.0351 | 0.0395 | 0.2270 |
| | $P$=30% | 0.0861 | 0.1405 | 0.1356 | 0.1210 | 0.3014 |
| Clevend | $P$=10% | 0.1074 | 0.2428 | 0.2420 | 0.2397 | 0.8679 |
| | $P$=20% | 0.0999 | 0.2575 | 0.2597 | 0.2666 | 0.7087 |
| | $P$=30% | 0.0943 | 0.2386 | 0.2670 | 0.1632 | 0.5343 |
| Wine | $P$=10% | 0.0979 | 0.3487 | 0.1724 | 0.1824 | 0.6765 |
| | $P$=20% | 0.1079 | 0.3156 | 0.1774 | 0.2693 | 0.5406 |
| | $P$=30% | 0.0960 | 0.2978 | 0.2727 | 0.2667 | 0.4986 |
| Abalone | $P$=10% | 0.0784 | 0.2938 | 0.2514 | 0.2421 | 0.8435 |
| | $P$=20% | 0.0851 | 0.2864 | 0.2896 | 0.2466 | 0.4600 |
| | $P$=30% | 0.0731 | 0.1764 | 0.2351 | 0.1501 | 0.6802 |
| Slump | $P$=10% | 0.0631 | 0.0538 | 0.0509 | 0.0503 | 0.1409 |
| | $P$=20% | 0.0539 | 0.0580 | 0.0570 | 0.0483 | 0.2408 |
| | $P$=30% | 0.0757 | 0.1892 | 0.1549 | 0.1608 | 0.3072 |
| Strike | $P$=10% | 0.0294 | 0.0556 | 0.0427 | 0.0523 | 0.2683 |
| | $P$=20% | 0.0270 | 0.0620 | 0.0684 | 0.0497 | 0.4574 |
| | $P$=30% | 0.1141 | 0.4769 | 0.4898 | 0.5277 | 0.8411 |

In order to fully verify the performance of the proposed method in the non-Gaussian noise environment, further experiments will be carried out in the α-stable noise environment next. The $\alpha$-stable noise is generated by the $\alpha$-stable distribution. The noise is directly added to the output $y$ of the training set samples, and its characteristic function can be simply expressed as $\psi(x)=\exp(-\rho\,|\,x\,|^{\mu})$ , where $\rho = 0.1, \mu = 1.5$ [31]. The test results of each variant on the corresponding data sets are recorded in Table 5, and the best results are marked in bold. Table 6 shows the time consumed by each variant in the corresponding training process. According to the experimental results in Tables 5 and 6, it can be obtained that:

(1) In the $\alpha$-stable noise environment, the performance of M-BLS is very unstable. In the test of multiple data sets, the test RMSE of M-BLS is slightly lower than that of traditional BLS only on Bodyfat and Wine data sets, and the performance decreases in other cases. For $\alpha$-stable noise, the preset loss function is difficult to directly match the noise distribution. Huber loss usually employs linear decay for large residuals, while impulsive residuals for $\alpha$-stable noise often require faster decay. Although Cauchy loss corresponds to a stable distribution with $\alpha$=1, its logarithmic decay rate is still slower than the extreme impulse demand of $\alpha$-stable noise, and the fixed scale parameter cannot adapt to the noise intensity change. Although the exponential decay of Welsch loss is suitable for extreme impulsive noise, it cannot dynamically adjust the decay rate according to the noise impulse strength as a fixed function. M-BLS based on fixed-form M-estimator function is almost invalid for impulsive noise environment.

(2) AR-BLS shows better performance than M-BLS on all datasets. Except on the Abalone dataset, where its test RMSE is on par with the basic BLS, AR-BLS can always achieve lower test RMSE than the basic BLS in all other cases. For $\alpha$-stable noise, the adaptive robust kernel can adjust the parameter $\alpha$ continuously in the iterative process of the algorithm, so that the shape of the loss function matches the impulsive characteristics of $\alpha$-stable noise. In the early iteration of the algorithm, the $\alpha$-stable noise has strong impulse characteristics, and the $\alpha$ of the adaptive robust kernel automatically decreases to negative value, the loss function approaches Welsch loss (exponential decay), and the weight of the extreme impulse residual approaches 0, which reduces the interference of noise on the model. After several iterations of the algorithm, the weight of the noise samples is continuously reduced, $\alpha$ is gradually approaching 2, and the loss function is approaching the L2 loss, which retains the fitting accuracy of the model for the interior points. This data-driven parameter adaptation method enables the loss function to adjust in real time with the noise pulse intensity, avoiding the accumulation of bias caused by the preset shape error of the fixed function, thus significantly enhancing the robustness of AR-BLS in dealing with $\alpha$-stable noise.

(3) Compared with the traditional BLS, the training time of each variant shows a certain degree of growth. Among them, the increase of M-BLS is relatively small, while AR-BLS requires more training time. However, in terms of absolute time, all variants consume very limited training time. Considering the performance improvement achieved by AR-BLS, the proposed method maintains the advantage of BLS fast training in general.

Table 5

Performance of different methods on regression datasets with $\alpha$-stable noise

| Dataset | The RMSE of test data (RMSE±STD $\times 10^{-2}$) | | | | |
|---|---|---|---|---|---|
| | BLS | M-BLS(Huber) | M-BLS(Cauchy) | M-BLS(Welsch) | AR-BLS |
| Housing | 13.44±3.03 | 13.45±3.35 | 13.52±3.41 | 13.67±3.33 | 13.37±3.52 |
| Bodyfat | 20.18±2.41 | 19.72±2.61 | 19.60±2.61 | 19.77±2.60 | 19.35±2.51 |
| Clevend | 91.65±1.84 | 91.01±2.00 | 90.96±1.86 | 91.24±1.90 | 90.56±2.19 |
| Wine | 30.30±0.55 | 30.19±0.64 | 30.07±0.62 | 30.30±0.64 | 29.23±0.68 |
| Abalone | 15.76±2.36 | 15.97±2.00 | 15.97±1.91 | 16.00±2.00 | 15.76±2.32 |
| Slump | 12.13±5.09 | 12.69±4.99 | 12.99±4.87 | 12.87±4.87 | 11.72±4.99 |
| Strike | 99.24±1.60 | 102.39±8.49 | 101.45±5.54 | 100.18±3.12 | 99.22±1.61 |

Table 6

Training time of different methods on regression datasets with $\alpha$-stable noise

| Dataset | The training time of different methods (s) | | | | |
|---|---|---|---|---|---|
| | BLS | M-BLS(Huber) | M-BLS(Cauchy) | M-BLS(Welsch) | AR-BLS |
| Housing | 0.0137 | 0.0334 | 0.0340 | 0.0331 | 0.1412 |
| Bodyfat | 0.0391 | 0.0997 | 0.0724 | 0.0598 | 0.2711 |
| Clevend | 0.0478 | 0.0782 | 0.0668 | 0.0708 | 0.4428 |
| Wine | 0.1374 | 0.7671 | 0.7574 | 0.8166 | 0.9442 |
| Abalone | 0.1365 | 0.6447 | 0.6567 | 0.6173 | 0.7140 |
| Slump | 0.304 | 0.0458 | 0.0481 | 0.0406 | 0.0703 |
| Strike | 0.0321 | 0.0515 | 0.0479 | 0.0550 | 0.0863 |

## 4.2 Experiments on practical application

As the most widely used building material in modern civil engineering, concrete plays an irreplaceable role in all kinds of engineering buildings such as roads, Bridges and high-rise buildings. As one of the core performance indicators of concrete, compressive strength not only determines the bearing capacity and safety of the structural system, but also affects the overall stability and service life of the engineering structure. However, the compressive strength of concrete is affected by the coupling effect of multiple factors, such as cement type, aggregate gradation, water-binder ratio, mineral admixture, curing conditions, etc. How to establish an efficient, accurate and widely applicable method for predicting the compressive strength of concrete according to the existing multi-source information such as concrete mix parameters, raw material characteristics and actual engineering environmental factors has become a key issue in the research of concrete materials.

In order to evaluate the practicability of the proposed method, 1030 different concrete mix ratios are collected in this subsection[32]. By systematically collating the collected data, nine key variables were extracted, eight of which were used as input variables, including the amount of cement, slag, fly ash, water, strong plastic agent, coarse aggregate, fine aggregate in concrete and the curing days. 1 is the output variable, which is the compressive strength of concrete. See Table 7 for the details of each input feature.

To fully validate the testing performance of the proposed method on concrete data, this section introduces machine learning models such as linear regression (LR), support vector machine (SVM), multilayer perceptron (MP), and backpropagation neural network (BPNN) for comparative experiments. Moreover, in order to optimize the construction of subsequent models, a systematic statistical analysis of the concrete data set is carried out, and the specific statistical indicators are detailed in Table 8. According to the results in the table, except for the plasticizer, there are significant differences between the most values and quartiles of the other characteristic parameters, and the standard deviation of some parameters is 50% higher than the mean, which indicates a high degree of dispersion of the data. Further analysis shows that the absolute value of skewness of slag, plasticizer and maintenance days is high ( $\geq 0.8$), which significantly deviates from the normal distribution reference, indicating that the data distribution has strong asymmetry. It is worth noting that the characteristics of maintenance days show obvious peak distribution characteristics: its kurtosis value is as high as 12.17 (the general normal distribution kurtosis is 3), reflecting the existence of super-normal aggregation in the mean area of the data. The above feature parameters show multiple complex characteristics.

Next, the concrete strength is predicted based on the proposed method. Firstly, the collected 1030 samples are divided, and the ratio of training set and test set is 7:3. In order to eliminate the influence of feature scale differences on the learning performance of the model, all samples are normalized. Table 9 lists the RMSE, MAE and the corresponding training time of the concrete strength prediction results of each method, where the best result is marked in bold. It is worth noting that since the model input is normalized, the actual prediction strength needs to be obtained by de-normalizing the model output.

Table 7

Introduction of concrete characteristic data

| Feature | The function of the concrete composition and the influence on the strength of concrete |
|---|---|
| Cement | Responsible for binding stone and aggregate, can improve the early strength of concrete, but excessive use may cause concrete to crack and reduce its long-term strength. |
| Slag | As a mineral admixture, it improves durability. The early strength is slightly lower than that of pure cement concrete, but the later strength (after 90 days) is higher. |
| Coal ash | As a mineral admixture, it improves fluidity. The early strength is significantly lower than that of pure cement concrete, but the later strength (after 90 days) may be higher. |
| Water | The key medium that constitutes concrete will directly affect the water-binder ratio. Under the premise of ensuring fluidity, the lower the water-binder ratio, the higher the strength of the concrete. |
| Plasticizer | As a water reducer, it improves the density of concrete. The strength of concrete can be enhanced by reducing the water-binder ratio, but excessive use may lead to uneven strength of the concrete. |
| Coarse aggregate | To form the concrete framework, excessive usage will lead to an increase in porosity and a decrease in concrete strength, while too low usage will increase the risk of cracking. |
| Fine aggregate | sed for filling the gaps of coarse aggregates. Excessive sand ratio will lead to an increase in the actual water-binder ratio of concrete and a decrease in compressive strength. |
| Maintenance days | The strength of concrete increases with the growth of curing time. The early strength (3 to 28 days) increases rapidly, while the later strength (after 28 days) gradually slows down. |

Table 8

Training time of different methods on regression datasets with different non-Gaussian noise

| Feature | Mean | Standard deviation | Minimum | 25% | 50% | 75% | Maximum | Skewness | Kurtosis value |
|---|---|---|---|---|---|---|---|---|---|
| Cement | 281.17 | 104.51 | 102.00 | 192.38 | 272.90 | 350.00 | 540.00 | 0.51 | -0.52 |
| Slag | 73.90 | 86.28 | 0.00 | 0.00 | 22.00 | 142.95 | 359.40 | 0.80 | -0.51 |
| Coal ash | 54.19 | 64.00 | 0.00 | 0.00 | 0.00 | 118.30 | 200.10 | 0.54 | -1.33 |
| Water | 181.57 | 21.35 | 121.80 | 164.90 | 185.00 | 192.00 | 247.00 | 0.07 | 0.12 |
| Plasticizer | 6.20 | 5.97 | 0.00 | 0.00 | 6.40 | 10.20 | 32.20 | 0.91 | 1.41 |
| Coarse aggregate | 972.92 | 77.75 | 801.00 | 932.00 | 968.00 | 1029.40 | 1145.00 | -0.04 | -0.60 |
| Fine aggregate | 773.58 | 80.18 | 594.00 | 730.95 | 779.50 | 824.00 | 992.60 | -0.25 | -0.10 |
| Maintenance days | 45.66 | 63.17 | 1.00 | 7.00 | 28.00 | 56.00 | 365.00 | 3.27 | 12.17 |

Tabel 9
Performance and training time of different methods for concrete strength prediction

| Algorithm | MAE±STD | Training time (s) | RMSE±STD | Training time (s) |
|---|---|---|---|---|
| LR | 11.30±1.54 | 0.0205 | 11.30±1.14 | 0.0010 |
| SVM | 9.42±1.23 | 9.7217 | 11.60±1.12 | 6.0104 |
| MP | 7.99±0.87 | 27.7583 | 9.61±0.82 | 10.6336 |
| BPNN | 8.07±1.03 | 5.1834 | 10.26±0.90 | 0.8562 |
| BLS | 8.00±0.96 | 0.0089 | 10.14±0.83 | 0.0161 |
| M-BLS(Huber) | 6.42±0.44 | 0.0492 | 8.30±0.51 | 0.0719 |
| M-BLS(Cauchy) | 6.30±0.40 | 0.0396 | 8.21±0.49 | 0.0388 |
| M-BLS(Welsch) | 6.32±0.42 | 0.0512 | 8.20±0.50 | 0.0416 |
| AR-BLS | 6.19±0.34 | 0.5806 | 8.19±0.49 | 0.6753 |

Tabel 10
Performance and training time of different methods for power prediction.

| Algorithm | | BLS | M-BLS(Huber) | M-BLS(Cauchy) | M-BLS(Welsch) | AR-BLS |
|---|---|---|---|---|---|---|
| Short term power prediction | RMSE±STD | 0.2743±0.0170 | 0.2419±0.0279 | 0.2443±0.0286 | 0.2497±0.0315 | 0.2389±0.0282 |
| | Training time (s) | 0.0021 | 0.0446 | 0.0867 | 0.0595 | 0.8521 |
| Ultra short term power prediction | RMSE±STD | 0.5019±0.0236 | 0.4166±0.1657 | 0.4122±0.1577 | 0.4194±0.1686 | 0.3669±0.1105 |
| | Training time (s) | 0.0798 | 0.7262 | 0.9806 | 0.7921 | 8.1120 |

From the experimental results in Table 9, it can be obtained that:

(1) Compared with BLS, the performance advantage of M-BLS on concrete strength prediction task is obvious, and the RMSE and MAE of the three M-BLS are lower than that of the traditional BLS. Among them, M-BLS(Cauchy) achieves the lowest RMSE and M-BLS(Welsch) achieves the lowest MAE. It can be seen that the M-estimator function that achieves the optimal performance is not fixed for different performance metrics. However, AR-BLS always achieves the best performance for both RMSE and MAE. The adaptive robust kernel can cover a variety of kernel morphologies and even generate intermediate states by continuously optimizing the shape parameters. This not only avoids the trial-and-error cost of manual parameter tuning in the application of fixed-form M-estimation functions, but also provides a unified optimization framework for complex scenarios.

(2) Compared with other variants, AR-BLS requires longer training time. However, from the perspective of overall performance, this sacrifice in training speed is acceptable. The proposed method can achieve good performance without manually adjusting parameters according to experience and multiple tests, which makes the model have high application value in dealing with practical problems.

## 5. Conclusion

This paper proposes a new variant of BLS based on adaptive robust kernel to improve the performance of conventional BLS under non-Gaussian noise. AR-BLS can automatically adjust its robustness under different noises through adaptive parameter optimization without additional manual parameter tuning or cross-experimental validation. Based on Zangwill's theorem, the iterative convergence of AR-BLS is proved. Simulation experiments on UCI regression data sets and concrete strength prediction tasks verify the effectiveness of the proposed method.

However, the focus of this paper is to explore the application of AR-BLS in regression tasks, and its applicability in classification tasks has not been explored. Therefore, in the future research planning, the proposed method can be considered to be extended to the field of classification tasks, and its potential and value in this regard can be deeply explored.

## Acknowledgments

This work was supported by the National Natural Science Foundation of China (NSF) 62171388.